\documentclass[notitlepage,english,aps,floats,onecolumn,showpacs,nofootinbib,floatfix]{revtex4-2}
\usepackage{pslatex}
\usepackage[T1]{fontenc}
\usepackage[latin1]{inputenc}
\usepackage{graphicx}
\usepackage{epsfig}
\usepackage{longtable}
\usepackage{float}
\usepackage{calc}
\usepackage{ifthen}
\usepackage{amsmath}
\usepackage{hyperref}
\usepackage{amssymb}

\usepackage{color}

{
{
{
\newcommand{\bea}{\begin{eqnarray}}
\newcommand{\eea}{\end{eqnarray}}

\newcommand{\nc}{\newcommand}
\nc{\renc}{\renewcommand}
\nc{\eqs}[2]{\mbox{Eqs.~(\ref{#1},\,\ref{#2})}}
\nc{\eq}[1]{\mbox{Eq.~(\ref{#1})}}
\nc{\figs}[2]{\mbox{Figs.~(\ref{#1},\,\ref{#2})}}
\nc{\fig}[1]{\mbox{Fig~.(\ref{#1})}}
\nc{\be}[1]{\begin{equation} \mbox{$\label{#1}$}}
\nc{\ee}{\vspace{0.1cm}\end{equation}}

\newcommand{\bean}{\begin{eqnarray*}}
\newcommand{\eean}{\end{eqnarray*}}

\def\GeV{{\rm \ GeV}}

\def\lae{\;^{<}_{\sim} \;} 

\def\gae{\; ^{>}_{\sim} \;}

\begin{document}

\title{Unitarity-Conserving Non-Minimally Coupled Inflation and the ACT spectral index} 

\author{John McDonald }
\email{j.mcdonald@lancaster.ac.uk}
\affiliation{Dept. of Physics,  
Lancaster University, Lancaster LA1 4YB, UK}

\begin{abstract} 

The Atacama Cosmology Telescope (ACT) collaboration has reported a scalar spectral index $ n_s~=~0.9743~\pm~0.0034 $. This is substantially larger than the classical prediction of non-minimally coupled inflation models such as Higgs Inflation, $n_s \approx 0.965$. Here we revisit the unitarity-conserving non-minimally coupled inflation model proposed in \cite{rl1}. We show that when the inflaton is a complex non-minimally coupled gauge singlet scalar with additional interactions in the Jordan frame to maintain unitarity, the model predicts $n_s = 0.9730$ and $r \approx 9 \times 10^{-6}$ for scalar self-coupling $\lambda = 0.1$. 
  
\end{abstract} 
 \pacs{}
 
\maketitle

\section{Introduction}

Inflation seeks to explain the observed homogeneity, isotropy and flatness of the Universe and the magnitude and scale dependence of the observed primordial density perturbations via an early era of exponential expansion. This is possible if the energy density of the Universe is dominated by a scalar potential $V(\phi)$ with the right form. The scale dependence of the primordial density perturbations is parameterised by the scalar spectral index, $n_s$. This has been observationally determined by observing the temperature fluctuations, polarisation and lensing of the CMB, combined with Baryon Acoustic Oscillations (BAO) from galaxy surveys. Planck \cite{planck} obtained a value $n_s = 0.9652 \pm 0.0042$ (TT, TE, EE + lowE + lensing), based on CMB temperature, polarisation, low multipole polarisation (lowE) and lensing, combining the results of two likelihood methods ({\it Plik} and {\it CamSpec}). Including BAO observations from 6dFGS, SDSS-MGS and BOSS DR12 increases the mean $n_s$ to around 0.967 \cite{planck}. 

A new determination of the scalar spectral index by the ACT collaboration \cite{act} finds $n_s = 0.9744 \pm 0.0034$. 
ACT is able to probe CMB polarisation up to $l = 1700$. This is larger than Planck, which is noise dominated in polarisation at $l > 800$. The new result for $n_s$ is obtained by combining the small scale CMB temperature and polarisation data from ACT DR6 with the large scale data from Planck (forming a joint P-ACT set) and combining this with CMB lensing data from ACT and Planck and with BAO data from DESI DR1, with $\Lambda$CDM assumed throughout.

It is important to ask what the implications of this new determination are for well-motivated inflation models. (For recent discussions of models in light of ACT, see \cite{actm1}-\cite{actm17}.) It has been noted that the ACT spectral index is not in good agreement with the theoretical result predicted by a range of inflation models, $n_{s} = 1-2/N_{*}$. In particular. Higgs Inflation \cite{bs} at the classical level which predicts $n_{s} = 0.965$ when $N_{*} = 57$. 

Higgs Inflation is one example of a general class of inflation models in which the inflaton can be a conventional scalar particle with couplings to itself and to other particles of strength typical of the Standard Model (SM) and its extensions \cite{salopek}. 
This opens up the attractive possibility that the inflation can be due to a conventional scalar particle of the SM and its extensions. We recently showed that Higgs Inflation can be in reasonable agreement with the ACT spectral index when the otherwise metastable Higgs potential is stabilised by the addition of vector-like quarks to the SM of mass of the order of a TeV
\cite{jmact,vlf1}. In this case the tensor-to-scalar is predicted to be large, $r \approx 0.01$. 

However, there is a concern that Higgs Inflation is fundamentally inconsistent due to unitarity violation from the non-minimal coupling \cite{uviol1,uviol2}. The non-minimal coupling causes Higgs scattering via graviton exchange to become strongly coupled at $E \sim M_{Pl}/\xi$. Whilst there is evidence that unitarity is conserved non-perturbatively and that Higgs Inflation is a self-consistent theory \cite{uv1,uv2,uv3}, it is still possible that the theory will break down at scattering energies greater than $M_{Pl}/\xi$. In this case the theory must be altered at higher energies to conserve unitarity, in which case it is not certain that the predictions of Higgs Inflation will remain valid.   

In a paper some time ago \cite{rl1}, a unitarity-conserving variant of non-minimally coupled inflation was proposed. The idea was that in the Jordan frame, the non-minimal coupling should always be accompanied by non-renormalisable terms that maintain unitarity. The minimal requirement for this is that the non-renormalisable terms which lead to unitarity violation should be removed in the Einstein frame, leaving canonical kinetic terms for the Higgs boson plus the modified Higgs potential. The predictions of the model, which we review here, were that $n_{s} = 1 - 3/(2 N_{*})$ and $r = 2/(\sqrt{\xi}N^{3/2})$. Using the data available at the time\footnote{The prediction in \cite{rl1} was before the Higgs boson experimental discovery and Planck CMB data.}, $n_{s}$ at the WMAP pivot scale, $k_{*} = 0.002 {\rm Mpc}^{-1}$, which corresponds to $N_{*} \approx 57$ assuming instantaneous reheating, was predicted to be $n_{s} = 0.974$, with $r$ predicted to be $r \approx 6 \times 10^{-6}$.

For the case of Higgs Inflation, the minimal unitarity-conserving model in \cite{rl1} works at the classical level, but because the Higgs is charged under the SM, the gauge covariant derivatives of the Higgs kinetic term in the Einstein frame add a gauge interaction which leads to a large quantum correction to the Higgs potential. In \cite{rl1} it was suggested that a SUSY version of the model might control the quantum corrections. However, an easier way to achieve a consistent model is simply to add a complex gauge singlet scalar to the SM to serve as the inflaton, in which case there are no problematic gauge interactions and quantum corrections. 

\section{The Inflaton Sector} 

In order to justify the additional terms in the Jordan frame that maintain unitarity conservation and modify the non-minimally coupled inflation model predictions, the unitarity-conserving model actually {\it requires} that unitarity is violated by the non-minimal coupling of the inflaton to the Ricci scalar. This requires that there is more than one real scalar with a large non-minimal coupling, since the model with only one real singlet scalar non-minimally coupled to gravity does not violate unitarity if the Einstein frame potential does not violate unitarity, and therefore the results of conventional non-minimally coupled inflation apply \cite{rl0,uviol2,rl2}. We will consider the simplest example where the inflaton is the real part of a non-minimally coupled complex scalar.  

For the complex inflaton $\Phi$, the conventional non-minimally coupled action in the Jordan frame is 
\be{e1} S_{J} = \int d^{4}x \sqrt{-g} \left( -\frac{M_{Pl}^{2} R}{2} - \xi \Phi^{\dagger} \Phi R + g^{\mu \nu} \partial_{\mu} \Phi^{\dagger} \partial_{\nu} \Phi - V(|\Phi|) \right) ~,\ee
where 
\be{e2} V(|\Phi|) = \lambda |\Phi|^4   ~.\ee 
When transformed to the Einstein frame via the conformal transformation $\tilde{g}_{\mu \nu} = \Omega^{2} g_{\mu \nu}$, 
the action becomes 
\be{e3}  S_{E} = \int d^{4}x 
 \sqrt{-\tilde{g} } 
\left(
 - \frac{M_{Pl}^{2}}{2} \tilde{R} + 
\frac{1}{\Omega^2}\tilde{g}^{\mu \nu} \partial_{\mu} \Phi^{\dagger} \partial_{\nu} \Phi 
+ \frac{3 \xi^{2}}{\Omega^{4}  M_{pl}^{2} } 
\tilde{g}^{\mu \nu} \partial_{\mu}(\Phi^{\dagger} \Phi) \partial_{\nu}(\Phi^{\dagger} \Phi) -V_{E}(|\Phi|) \right)
~,\ee 
where 
\be{e4} \Omega^2 = 1 + \frac{2 \xi |\Phi|^{2}}{M_{Pl}^2}  ~\ee
and 
\be{e5} V_{E}(|\Phi|) = \frac{V(|\Phi|)}{\Omega^{4}}   ~.\ee 
Tree-level unitarity violation due to graviton-mediated Higgs scattering in the Jordan frame manifests itself in the Einstein frame via the $\xi$ dependent interactions from the second and third terms of $S_{E}$. (We argue below that the modified potential $V_{E}(|\Phi|)$ does not introduce any unitarity violation.) If this indicates true unitarity violation at scattering energies $E \gae M_{Pl}/\xi$ (in the case of the third term) or at $ E \gae M_{Pl}\sqrt{\xi}$ (in the case of the second term), then the theory is inconsistent and must be modified. 

This can be done by adding terms to the Jordan frame action which cancel the unitarity violation. The non-minimal coupling can be thought of as part of the complete Planck scale theory that has had its scale lowered from $M_{Pl}$. To maintain unitarity, the non-minimal kinetic term must be accompanied by additional terms that form a complete unitarity-conserving unit. The minimal requirement is that these should eliminate the $\xi$ dependent terms in the Einstein frame except for the $\xi$ dependence of the potential. The Einstein frame action is then 
\be{e6} S_{E} = \int d^{4} x 
\left(
 -\frac{M_{Pl}^{2} \tilde{R}}{2} 
 + \tilde{g}^{\mu \nu} \partial^{\mu} \Phi^{\dagger} \partial^{\nu} \Phi - \frac{V(|\Phi|)}{\Omega^{4}}   \right) ~.\ee 
The Jordan frame action which achieves this is 
\be{e7}  S_{J} = \int d^{4} x \sqrt{-g} \left( -\frac{M_{P}^{2} R}{2} - \left\{\xi \Phi^{\dagger} \Phi R - \frac{2 \xi \Phi^{\dagger} \Phi}{M_{Pl}^{2}} g^{\mu \nu} \partial_{\mu} \Phi^{\dagger} \partial_{\nu} \Phi + \frac{3 \xi^{2}}{\Omega^{2} M_{Pl}^{2}} g^{\mu \nu} \partial_{\mu}\left(\Phi^{\dagger}\Phi\right) \partial_{\nu}\left(\Phi^{\dagger}\Phi\right) \right\}  + g^{\mu \nu} \partial_{\mu} \Phi^{\dagger} \partial^{\nu} \Phi   - V(|\Phi|)  \right) ~.\ee
The terms in the curly brackets form the unitarity-conserving unit in the Jordan frame.

\section{Unitarity}

For inflation to be possible with a conventionally large self-coupling for $\Phi$, say $\lambda \sim 0.1$, it is essential to retain the modified Einstein frame potential. A concern is that the non-polynomial potential will lead to dangerous unitarity violation. In this section we argue that this is not the case.

To estimate the magnitude of the scattering energy at unitarity violation we use the optical theorem, which follows from unitarity conservation. This relates the total amplitude for  2 $\rightarrow$ anything scattering to the $2 \rightarrow 2$ elastic scattering amplitude in the forward direction. Dimensionally, the inelastic scattering amplitude from non-renormalisable terms of the form $\lambda {\cal O}(\phi)/\Lambda^{n}$, where ${\cal O}(\phi)$ is an operator composed of $\phi$ and possibly its derivatives, is ${\cal M} \sim \lambda E^{n}/\Lambda^{n}$, and the cross-section is 
\be{ub1} \sigma_{inelastic} \sim \frac{|{\cal M}|^{2}}{E^{2}}  \sim \frac{\lambda^{2} E^{2n - 2}}{\Lambda^{2n}} ~.\ee  
The unitarity bound from the optical theorem requires that the total inelastic scattering cross-section $\sigma_{TOT}$  satisfies 
\be{ub2}  \sigma_{TOT} \lae \frac{4 \pi}{E^{2}}   ~,\ee 
which follows from the upper bound on the elastic scattering partial wave amplitude from unitarity, under the assumption  that the elastic scattering amplitude is dominated by low multipoles \cite{rl2}.  
Therefore the bound on the leading inelastic contribution to $\sigma_{TOT}$, which typically corresponds to the lowest dimension non-renormalisable operator (usually $n = 5$ or $n = 6$), is 
\be{ub3} E \lae \frac{\Lambda}{\lambda^{1/n}}  ~.\ee 

\subsection{Unitarity violation due to the kinetic term} 

In the original non-minimally coupled inflation model, there are three possible sources of unitarity violation in the Einstein frame: (i) the kinetic term of the scalar, (ii) the derivative interaction term, and (iii) the Einstein frame potential. The derivative interaction term unambiguously violates unitarity at $E \sim M_{Pl}/\xi$, as can be seen from its coefficient, $3 \xi^{2}/(\Omega^{2} M_{Pl}^{2})$ in the limit $\Omega = 1$, and so must be removed. By 'unambiguous' we mean that it is not dependent upon an expansion of the conformal factor $\Omega$. 

For the kinetic term, on the other hand, we need to expand the conformal factor to be able to quantise the theory. We can state the kinetic term as  
\be{u1} {\cal L}_{k} =
 \frac{\left( \partial_{\mu} \phi_{1} \partial^{\mu} \phi_{1} +  \partial_{\mu} \phi_{2} \partial^{\mu} \phi_{2} \right) }{2\left(1 + 
\frac{\xi}{M_{Pl}^{2}}
\left(\phi_{1}^{2} + \phi_{2}^{2}\right)
\right) }   ~,\ee
where $\Phi = (\phi_{1} + i \phi_{2})/\sqrt{2}$. If we assume that the magnitudes of the $\phi_{i}$ ($i = 1,\,2$) dependent terms in the denominator are small compared to one, then we can simply Taylor expand the kinetic term as 
\be{u2}  {\cal L}_{k} \approx \frac{1}{2} 
\left( \partial_{\mu} \phi_{1} \partial^{\mu} \phi_{1} +  \partial_{\mu} \phi_{2} \partial^{\mu} \phi_{2} \right)   
\left(1 - \frac{\xi}{M_{Pl}^{2}}\left(\phi_{1}^{2} + \phi_{2}^{2}\right)  + ... \right)  ~.\ee
The interaction term
\be{u3}   - \frac{\xi}{2 M_{Pl}^{2}}
\left( \partial_{\mu} \phi_{1} \partial^{\mu} \phi_{1} +  \partial_{\mu} \phi_{2} \partial^{\mu} \phi_{2} \right)   
\left(\phi_{1}^{2} + \phi_{2}^{2}\right)   ~\ee
will violate unitarity at scattering energy $E \sim M_{Pl}/\sqrt{\xi}$. 

This is true for the vacuum with $\phi_{i} = 0$ and $\Omega = 1$. A related question is unitarity violation in the presence of a large background inflaton field. We define $\phi_{1,\,0}$ to be the background inflaton field and $\Phi = \frac{1}{\sqrt{2}} \left(\phi_{1,\,0} + \tilde{\phi}_{1} + i \tilde{\phi}_{2} \right)$. The non-minimal kinetic term is then 
\be{u4} {\cal L}_{k} = \frac{  \left( \partial_{\mu} \tilde{\phi}_{1} \partial^{\mu} \tilde{\phi}_{1} +  \partial_{\mu} \tilde{\phi}_{2} \partial^{\mu} \tilde{\phi}_{2} \right)}
{2 \left(1 + \frac{\xi}{M_{Pl}^{2}} \left(\phi_{1,\,0}^{2} + 2 \phi_{1,\,0} \tilde{\phi}_{1} + \tilde{\phi}_{1}^{2} + \tilde{\phi}_{2}^{2} \right) \right) }   ~.\ee
We assume that $\phi_{1,0} \gg M_{Pl}/\sqrt{\xi}$, in which case the non-minimal kinetic term becomes 
\be{u5} {\cal L}_{k} \approx \frac{M_{Pl}^{2}}{2 \xi \phi_{1,\,0}^{2}} \left( \partial_{\mu} \tilde{\phi}_{1} \partial^{\mu} \tilde{\phi}_{1} +  \partial_{\mu} \tilde{\phi}_{2} \partial^{\mu} \tilde{\phi}_{2} \right) 
\frac{1}{\left(1 + \frac{2 \tilde{\phi}_{1}}{\phi_{1,\,0}}  + \frac{\tilde{\phi}_{1}^{2}}{\phi_{1,\,0}^{2}}  + \frac{\tilde{\phi}_{2}^{2}}{\phi_{1,\,0}^{2}} \right)}
 ~.\ee 
We can define canonically normalised scalars in the inflaton background as $\sigma_{i} = M_{Pl} \tilde{\phi}_{i}/(\sqrt{\xi} \phi_{1,\,0}) $. Then
\be{u6} {\cal L}_{k} \approx \frac{1}{2} \left( \partial_{\mu} \sigma_{1} \partial^{\mu} \sigma_{1} +  \partial_{\mu} \sigma_{2} \partial^{\mu} \sigma_{2} \right) \frac{1}{\left(1  + 
 \frac{2 \sqrt{\xi}}{M_{Pl}} \sigma_{1} + \frac{\xi}{M_{Pl}^{2}}  \left(\sigma_{1}^{2} + \sigma_{2}^{2} \right) \right)}   ~.\ee 
Assuming that $\sigma_{i}$ can be considered small compared to $M_{Pl}/\sqrt{\xi}$, this can be expanded as 
\be{u7}   {\cal L}_{k} \approx \frac{1}{2} \left( \partial_{\mu} \sigma_{1} \partial^{\mu} \sigma_{1} +  \partial_{\mu} \sigma_{2} \partial^{\mu} \sigma_{2} \right)\left(1  -
 \frac{2 \sqrt{\xi}}{M_{Pl}} \sigma_{1} - \frac{\xi}{M_{Pl}^{2}}\left(\sigma_{1}^{2} + \sigma_{2}^{2} \right) + ...\right) ~,\ee
The interaction terms will violate unitarity at $\tilde{E} \sim M_{Pl}/\sqrt{\xi}$, where $\tilde{E}$ is the energy in the Einstein frame. Therefore, both in the vacuum and in the presence of an inflaton background, unitarity is violated at $\tilde{E} \sim M_{Pl}/\sqrt{\xi}$. 

However, this result is reliant upon the assumption that $\sigma_{i}$ is small compared to $M_{Pl}/\sqrt{\xi}$, so that the non-minimal kinetic term can be Taylor expanded. A natural estimate of the magnitude of the canonically normalised $\sigma_{i}$ field in scattering processes at energy $\tilde{E}$ is $\sigma_{i} \sim \tilde{E}$. (In \cite{pu1}, we argued that this can be understood from the scattering of wavepackets of energy $\tilde{E}$ rather than plane waves. It can also be understood from the non-renormalisable interactions in the expansion acquiring factors of $\tilde{E}$ to form dimensionless quantities when converting into scattering amplitudes.) However, at $\tilde{E} \sim \Lambda = M_{Pl}/\sqrt{\xi}$ this implies that $\sigma_{i}$ cannot be considered small compared to $M_{Pl}/\sqrt{\xi}$ and so the Taylor expansion and calculation breaks down. In this case it is not certain that unitarity is violated by the kinetic term. Here we will take the conservative view that the estimate of the energy of unitarity violation remains valid in the limit where the expansion of the non-canonical kinetic term is just breaking down i.e. as $\tilde{E} \rightarrow M_{Pl}/\sqrt{\xi}$. In this case we must convert the non-canonical kinetic term to a canonical kinetic term to conserve unitarity. 

\subsection{The Einstein frame potential} 

We next consider the question of unitarity violation due to $V_{E}(|\Phi|)$, applying the same principle that the estimate of the energy of unitarity violation remains valid in the limit where the Taylor expansions are just breaking down. 

We first consider the case of a background inflaton field $\phi_{1,\,0} > M_{Pl}/\sqrt{\xi}$. This background can represent the inflaton during inflation. It can also be used to estimate scattering rates when $\tilde{E} > M_{Pl}/\sqrt{\xi}$, by taking the limit where the background field is of the same magnitude as the scalar field scattering at energy $\tilde{E}$, $\phi_{1,\,0} \sim \tilde{E}$.

We assume that the $\Phi$ kinetic term has already been converted to a canonical kinetic term in order to conserve unitarity, so that the $\phi_{i}$ are canonically normalised scalars. The potential is
\be{u8} V_{E} = \frac{\lambda |\Phi|^{4}}{\left( 1 + \frac{2 \xi |\Phi|^{2}}{M_{Pl}^{2}} \right)^{2} }  = \frac{\lambda M_{Pl}^{4}}{4 \xi^{2}}  
\frac{1}{ 
\left( 1 + \frac{M_{Pl}^{2}}{2 \xi |\Phi|^{2}} \right)^{2} }  ~.\ee
This can be written as
\be{u10} V_{E} = \frac{\lambda M_{Pl}^{4}}{4 \xi^{2}}
 \left(1 + \frac{M_{Pl}^{2}}{\xi \phi_{1,\,0}^{2} \left( 1 + 2 \frac{\tilde{\phi}_{1}}{\phi_{1,\,0}} + \frac{\tilde{\phi}_{1}^{2}}{\phi_{1,\,0}^{2}} + \frac{\tilde{\phi}_{2}^{2}}{\phi_{1,\,0}^{2}} \right) } \right)^{-2}  ~.\ee
Assuming that $\phi_{1,\,0}^{2} \gg M_{Pl}^{2}/\xi$, to leading order in $M_{Pl}^{2}/(\xi \phi_{1,\,0}^{2}$ we can expand this as 
\be{u11} V_{E} \approx \frac{\lambda M_{Pl}^{4}}{4 \xi^{2}}  \left(1 - \frac{2 M_{Pl}^{2}}{\xi \phi_{1,\,0}^{2}} \frac{1}{\left( 1 + 2 \frac{\tilde{\phi}_{1}}{\phi_{1,\,0}} + \frac{\tilde{\phi}_{1}^{2}}{\phi_{1,\,0}^{2}} + \frac{\tilde{\phi}_{2}^{2}}{\phi_{1,\,0}^{2}} \right)} \right) ~.\ee
We can estimate the magnitude of the non-renormalisable interactions by considering only $\tilde{\phi}_{1}$, since the non-renormalisable terms involving products of $\tilde{\phi}_{1}$ and $\tilde{\phi}_{2}$ will have a similar mass scale. Assuming that $\tilde{\phi}_{1}$ is small  compared to $\phi_{1,\,0}$, we can then Taylor expand the potential as
\be{u12} V_{E} = \frac{\lambda M_{Pl}^{4}}{4 \xi^{2}}\left( 1 + \frac{2 M_{Pl}^{2}}{\xi \phi_{1,\,0}^{2}} \sum_{n = 0}^{\infty} \left(-1\right)^{n+1} (n + 1)\left(\frac{\tilde{\phi}_{1}}{\phi_{1,\,0} } \right)^{n} \right) ~.\ee
The n-th term can be written as 
\be{u13} \left(-1\right)^{n + 1} \frac{\tilde{\phi}_{1}^{n}}{\Lambda_{n}^{n-4}}  ~\ee
where
\be{u14} \Lambda_{n} = \kappa_{n} \left(\frac{\xi^{3} \phi_{1,\,0}^{n + 2}}{\lambda M_{Pl}^{6}} \right)^{\frac{1}{n -4}} = \frac{\kappa_{n} M_{Pl}}{\sqrt{\xi} \lambda^{\frac{1}{n -4}}}  \left( \frac{\sqrt{\xi} \phi_{1,\,0}}{M_{Pl}} \right)^{\frac{n + 2}{n - 4} }~\ee
with
\be{u15} \kappa_{n} = \left( \frac{2}{n + 1} \right)^{\frac{1}{n - 4}}  ~.\ee
We find that the scale of the leading-order non-renormalisable operator, corresponding to the $\tilde{\phi}_{1}^{5}$ term, is 
\be{u14a}  \Lambda_{5} = \frac{M_{Pl}}{3 \sqrt{\xi} \lambda} \left(\frac{\sqrt{\xi} \phi_{1,\,0}}{M_{Pl}} \right)^{7}   ~\ee 
Since $\sqrt{\xi} \phi_{1,\,0} \gg M_{Pl}$, it follows that $\Lambda_{5} \gg M_{Pl}$ and therefore there is no dangerous unitarity violation from the Einstein frame potential.  

However, this assumes that $\tilde{\phi}_{1} < \phi_{1,\,0}$, so that the Taylor expansion is valid. Since $\tilde{\phi}_{1}$ is a canonically normalised field, we expect $\tilde{\phi}_{1} \sim \tilde{E}$ in scattering processes in the inflaton background. Therefore, for the calculation to be valid, we require that $\tilde{E} < \phi_{1,\,0}$. 
As before, 
we will conservatively assume that the estimate of the unitarity violation scale made when $\phi_{1,\,0} > \tilde{E}$ is still valid as $\phi_{1,0} \rightarrow \tilde{E}$ and the Taylor expansion is just breaking down. In this case, with $\phi_{1,\,0} = \tilde{E}$ we obtain 
\be{u15a} \Lambda_{5}(\tilde{E}) = \frac{\xi^{3} \tilde{E}^{7}}{3 \lambda M_{Pl}^{6}} ~.\ee 
Therefore the condition that $\tilde{E} \lae \Lambda_{5}(\tilde{E})$ requires that
\be{u16} \tilde{E} \gae \left(3 \lambda\right)^{\frac{1}{6} } \frac{M_{Pl}}{\sqrt{\xi}}   ~.\ee
This is true for $\tilde{E} = \phi_{1,\,0} > M_{pl}/\sqrt{\xi}$.

Therefore in scattering due to the potential at energies greater than $M_{Pl}\sqrt{\xi}$ unitarity is conserved. Unitarity is also conserved at $\tilde{E} \lae M_{Pl}/\sqrt{\xi}$, since in that case we can work without a background field and simply expand $\Omega^4$, giving $\Lambda = M_{Pl}/\sqrt{\xi}$. It follows that unitarity is likely to be conserved in scattering mediated by the Einstein frame potential at all energies and therefore there is no need to modify it.

\section{Interaction with SM Higgs}

So far we have considered only the complex scalar inflaton. In order to reheat the Universe, this should interact with SM particles. Moreover, we are interested in using a complex singlet scalar that can be a component of extensions of the SM. Such scalars can have important phenomenological effects, in particular for first-order electroweak phase transitions which may explain baryogenesis and may generate observable gravitational waves. We will therefore consider the case of the portal interaction with the Higgs boson. For the case of Einstein frame renormalisation (Prescription I renormalisation \cite{presc1,presc2}), we will show that the quantum corrections are suppressed during inflation and that the classical predictions for inflation observables apply. 

The Higgs kinetic term in the Einstein frame in the case of conventional non-minimally coupled inflation is
\be{u17} \frac{ (D_{\mu} H)^{\dagger} D^{\mu} H }{\left(1 + \frac{2 |\Phi|^{2}}{M_{Pl}^{2}} \right)}    ~.\ee 
This will cause unitarity violation as in the case of the $\Phi$ kinetic term and therefore must be modified to a canonical kinetic term. The additional term in the Jordan frame Lagrangian to achieve this is  
\be{u18} \frac{2 \xi |\Phi|^{2}(D_{\mu} H)^{\dagger} D^{\mu} H }{M_{Pl}^{2}}  ~.\ee 

The portal interaction in the Jordan frame is
\be{u19} \lambda_{\Phi\, H} |\Phi|^{2} |H|^{2}   ~.\ee 
In the Einstein frame this becomes 
\be{u20} \frac{   \lambda_{\Phi\, H} |\Phi|^{2} |H|^{2} }{\Omega^{4}}   ~.\ee

The frames in which the quantum corrections are renormalised corresponds to different UV completions of the model, with quite different predictions for the inflation observables \cite{uvframe}. The quantum correction to the inflaton potential for the case of renormalisation in the Jordan frame (Prescription II renormalisation) is complicated by the additional non-renormalisable interactions which conserve unitarity. However, if the portal coupling $\lambda_{\Phi H}$ and $\Phi$ self-coupling $\lambda$ are small compared to 1, we would expect the quantum corrections to the Jordan frame potential to be negligible and so the classical predictions for inflation observables to apply. On the other hand, in the case of Prescription I renormalisation, since the model is the Standard Model except for the modified potential, we can use the SM RG equations at renormalisation scale $\mu < M_{Pl}/\sqrt{\xi}$ to run the couplings up to a scale $\mu$ slightly less than $M_{Pl}/\sqrt{\xi}$, and then apply the 1-loop Coleman-Weinberg correction to obtain the quantum correction to the Higgs potential at $\phi > M_{Pl}/\sqrt{\xi}$. The 1-loop correction is 
\be{u21} \Delta V_{1-loop} = \sum_{i} \frac{C_{i} M_{i}^{4}}{64 \pi^2} \left[ \ln\left( \frac{M_{i}^{2}}{\mu^{2}} - K_{i} \right)  \right] ~,\ee 
where $(C_{i}, K_{i}) = (1, 3/2)$ for each real scalar of the Higgs doublet. The Higgs doublet scalar masses are
\be{u24} m_{H}^{2} = \frac{\lambda_{\Phi H} |\Phi|^{2}}{ 
\Omega^{4}}   ~.\ee 
The inflaton potential including the 1-loop correction from the portal coupling is then
\be{u25}   V_{E} = \frac{1}{\Omega^{4}} \left( \frac{\lambda \phi^{4}}{4} + \frac{\lambda_{\Phi H}^{2} \phi^{4}}{64 \pi^{2} \Omega^{4}} \ln \left( \frac{\lambda_{\Phi H} \phi^{2}}{2 \Omega^{4} \mu^{2}} \right) \right)~.\ee
At $\phi > M_{Pl}/\sqrt{\xi}$ this becomes 
\be{u25a}   V_{E} = \frac{1}{\Omega^{4}} \left( \frac{\lambda \phi^{4}}{4} + \frac{\lambda_{\Phi H}^{2} M_{Pl}^{4}}{64 \pi^{2} \xi^{2}} \ln \left( \frac{\lambda_{\Phi H} M_{Pl}^{4}}{2 \xi^{2}  \phi^{2} \mu^{2}} \right) \right) ~.\ee
Therefore the quantum correction becomes small relative to the tree-level potential when $\phi > M_{Pl}/\sqrt{\xi}$. Similarly, the 1-loop corrections from the $\Phi$ self-coupling will be small relative to the tree-level potential. Therefore the inflation predictions of the model for Prescription I renormalisation will be essentially the same as the classical predictions.

\section{Inflation predictions} 

The inflation predictions were first derived in \cite{rl1}. For the inflaton potential 
\be{v1} V_{E}(\phi) = \frac{\lambda \phi^{4}}{4 \left(1 + \frac{\xi \phi^{2}}{M_{Pl}^{2}} \right)^{2}} ~,\ee 
the number of e-foldings is 
\be{v2} N = \frac{\xi \phi^{4}}{16 M_{Pl}^{4}}   ~.\ee 
The slow-roll parameters are, to next-to-leading order for $\eta$ and to leading order for $\epsilon$, 
\be{v3} \eta = M_{Pl}^{2}\left( \frac{1}{V_{E}}\frac{d^{2} V_{E}}{d \phi^2}\right) = -\frac{12 M_{Pl}^{4}}{\xi \phi^4} + \frac{36 M_{Pl}^{6}}{\xi^{2} \phi^{6}} ~\ee
and
\be{v4} \epsilon = \frac{M_{Pl}^{2}}{2} \left( \frac{1}{V_{E}} \frac{d V_{E}}{d \phi} \right)^{2} = \frac{8 M_{Pl}^{6}}{\xi^{2} \phi^6} ~.\ee
The scalar spectral index is then 
\be{v5} n_{s} = 1 + 2 \eta - 6 \epsilon = 1 - \frac{3}{2 N} + O\left(\frac{1}{N^{3/2} \sqrt{\xi}}\right) ~,\ee
where the last term is generally negligible.  
The tensor-to-scalar ratio is 
\be{v6} r = 16 \epsilon = \frac{2}{\sqrt{\xi} N^{3/2} }  ~.\ee 
The curvature perturbation power spectrum is 
\be{v7} P_{ \zeta} =  \frac{V_{E}}{24 \pi^{2} \epsilon M_{Pl}^{4} } = \frac{\lambda N^{3/2}}{12 \pi^{2} \xi^{3/2}}  ~.\ee
The amplitude of the power spectrum,  $A_{s}$, denotes the value of $P_{\zeta}$ at the number of e-foldings corresponding to the CMB pivot scale, $N_{*}$.

\subsection{$N_{*}$ for instantaneous reheating} 

We assume instantaneous reheating via decay of the inflaton condensate to Higgs bosons. The number of e-foldings is determined from the time of horizon exit of the pivot scale wavenumber, corresponding to 
\be{rh1}  \left(\frac{a_{0}}{a}\right) \,k_{*} = 2 \pi H  ~,\ee
where $a$ is the scale factor '0' denotes present value.
During inflation $\rho \approx V_{0} = \lambda M_{Pl}^{4}/(4 \xi^{2})$, therefore 
\be{rh2} H = \left(\frac{V_{0}}{3 M_{Pl}^{2}} \right)^{1/2} = \left(\frac{\lambda}{12}\right)^{1/2} \frac{M_{Pl}}{\xi}   ~.\ee
$V_{0}$ is assumed to convert to radiation at the end of inflation, therefore the reheating temperature is
\be{rh3} T_{R} = \left(\frac{30 V_{0}}{\pi^{2} g(T_{R})}\right)^{1/4}  ~,\ee
where $T_{0}$ is the present CMB temperature and $g(T)$ is the effective number of relativistic degrees of freedom.  
Thus the LHS of \eq{rh1} is 
\be{rh4} \left(\frac{a_{0}}{a}\right) \, k_{*} = \left(\frac{a_{0}}{a_R}\right) \left(\frac{a_{R}}{a}\right) k_{*} = \left(\frac{g(T_{R})}{g(T_{0})} \right)^{1/3} \left(\frac{T_{R}}{T_{0}}\right)\, e^{N_{*}}\, k_{*}  ~.\ee
From \eq{rh1}, \eq{rh2} and \eq{rh3} we obtain 
\be{rh5} e^{N_{*}} = \frac{2 \pi T_{0}}{\xi^{1/2} k_{*}} \left(\frac{g(T_{0})}{g(T_{R})^{1/4}} \right)^{1/3} \left(\frac{\pi^{2}}{90} \right)^{1/4} \left(\frac{\lambda}{12}\right)^{1/4}   ~.\ee
We use \eq{v7} to write $\xi$ in terms of $\lambda$,
\be{rh6} \xi = \frac{\lambda^{2/3} N_{*}}{\left(12 \pi^{2} P_{\xi}\right)^{2/3} }   ~.\ee
\eq{rh5} then gives 
\be{rh7} e^{N_{*}} = \frac{c_{R} T_{0} P_{\xi}^{1/3}}{k_{*}}  \left(\frac{g(T_{0})}{g(T_{R})^{1/4}} \right)^{1/3} \frac{1}{\lambda^{1/12} N_{*}^{1/2}}   ~\ee
where 
\be{rh8} c_{R} = 2 \pi^{5/3}(12)^{\frac{1}{12}} \left(\frac{\pi^{2}}{90}\right)^{1/4} = 9.53  ~.\ee

\subsection{Predictions for $n_{s}$ and $r$} 

Using the Planck pivot scale, $k_{*} = 0.05 \,{\rm Mpc}^{-1} \equiv 3.20 \times 10^{-40} \GeV$, $g(T_{0}) = 3.91$, $g(T_{R}) = 106.75$, $A_s = 2.1 \times 10^{-9}$ and $T_{0} = 2.37 \times 10^{-13} \GeV$, we obtain
\be{v8b} N_{*} = 57.42 - \frac{1}{2} \ln N_{*} - \frac{1}{12} \ln \lambda  ~.\ee 
We find that $N_{*}$ is not very sensitive to $\lambda$; with $\lambda = 0.1$  we find $N_{*} = 55.60$ and with $\lambda = 10^{-4}$ we obtain $N_{*} = 56.18$. In Table 1 we show the inflation predictions as a function of $\lambda$. We find that $n_{s} = 0.9730$ to 0.9733 for $\lambda$ in the range $10^{-4}$ to 0.1, which is in good agreement with the ACT value. The tensor-to-scalar ratio is in the range $9 \times 10^{-6}$ to $9 \times 10^{-5}$, which is too small to be observed by the next generation of CMB observations.

\begin{table}[htbp]
\begin{center}
\begin{tabular}{ |c|c|c|c|c| }
\hline
$\lambda$ & $\xi$ & $N_{*}$ & $n_{s}$ & $r$    
\\
\hline
0.1
& $3.04 \times 10^{5}$ 
& $55.60$ & 0.9730 & $8.75 \times 10^{-6}$  
\\
0.01
& $6.57 \times 10^{4}$ 
& $55.79$ & 0.9731 & $1.87\times 10^{-5}$  
\\
$10^{-4}$
& $3.08 \times 10^{3}$ 
& $56.18$ & 0.9733 & $8.55 \times 10^{-5}$  
\\
\hline
\end{tabular}
\caption{Non-minimal coupling and inflation predictions as a function of $\lambda$.}
\end{center}
\end{table}

\section{Conclusions} The unitarity-conserving non-minimally coupled inflation model, originally proposed in \cite{rl1}, predicts $n_{s} = 0.9730$ and $r \approx 9 \times 10^{-6}$ for scalar self-coupling $\lambda = 0.1$, for the case of renormalisation of the quantum corrections in the Einstein frame and instantaneous reheating. This is in very good agreement with the recent ACT value, whilst the tensor-to-scalar ratio is too small to be observed by next-generation CMB experiments which will probe down to $r \sim 10^{-3}$. The results are insensitive to the scalar self-coupling, with $n_{s} = 0.9733$ for $\lambda = 10^{-4}$. The model requires that the inflaton is a complex gauge singlet scalar, in order to evade the large quantum corrections that would otherwise generally arise from the interaction of the inflation with the gauge bosons. The model offers the possibility of unifying a complex singlet scalar extension of the SM, which may play a role in enabling a first-order electroweak phase transition for electroweak baryogenesis and the production of primordial gravitational waves, with the inflaton.

\end{document}